\documentclass{elsart}
\journal{Physica A}

\usepackage{amssymb,amsfonts,graphicx}
\usepackage{graphicx}
\usepackage{dcolumn}
\usepackage{bm}

\begin{document}

\begin{frontmatter}

\title{Statistical Analysis of the Metropolitan Seoul Subway System: Network Structure and Passenger Flows}

\author[swu2,bu2]{Keumsook Lee}
\author[bu1]{Woo-Sung Jung}
\author[swu1]{Jong Soo Park}
\author[bu1,snu]{M. Y. Choi\corauthref{cor1}}
\corauth[cor1]{Permanent address: Department of Physics and Astronomy, Seoul National University, Republic of Korea}

\address[swu2]{Department of Geography, Sungshin Women's University, Seoul 136-742, Republic of Korea}
\address[bu2]{Center for Transportation Studies, Boston University, Boston, MA 02215, USA}
\address[bu1]{Center for Polymer Studies and Department of Physics, Boston University, Boston, MA 02215, USA}
\address[swu1]{School of Computer Science and Engineering, Sungshin Women's University, Seoul 136-742, Republic of Korea}
\address[snu]{Department of Physics and Astronomy, Seoul National University,
Seoul 151-747, Republic of Korea}

\begin{abstract}
The Metropolitan Seoul Subway system, consisting of 380 stations, provides the major transportation mode in the metropolitan Seoul area. Focusing on the network structure, we analyze statistical properties and topological consequences of the subway system. We further study the passenger flows on the system, and find that the flow weight distribution exhibits a power-law behavior. In addition, the degree distribution of the spanning tree of the flows also follows a power law.
\end{abstract}


\end{frontmatter}

Complex networks have been an active research topic in the physics
community since models for complex networks were announced
\cite{watts98,barabasi99}. Subsequently, numerous real networks
observed in biological and social systems as well as physical ones
have been studied
\cite{barabasi99-2,newman03,jeong03,amaral00,barrat04,guimera05,bagler04,li04,latora02,seaton04,sienkiewicz05,jung08,brockman06}.
Those studied also include transportation systems such as airline
networks, subway networks, and highway systems. For example,
studies of world-wide airport networks as well as Indian and
Chinese airport networks have disclosed small-world behaviors and
truncated power-law distributions
\cite{amaral00,barrat04,guimera05,bagler04,li04}. For the subway
systems in Boston and Vienna, various network properties, such as
the clustering coefficient and network size, have been reported
\cite{latora02,seaton04}. Further, statistical properties of the
Polish public transport network have been examined
\cite{sienkiewicz05} and the Korean highway system has been
analyzed with respect to the gravity model \cite{jung08}.

In this manuscript, we consider the Metropolitan Seoul Subway
(MSS) system, which consists of $N=380$ stations, and serves as
the major public transportation mode in the greater Seoul area,
Republic of Korea. The system in an earlier phase was analyzed
with regard to the accessibility measurement \cite{lee98}. In the
present phase, the maximum distance between a pair of stations in
the system is $126\,\textrm{km}$ while the minimum value is
$238\,\textrm{m}$. When we construct the subway network with $N$
nodes, each corresponding to a station, the number of links
connecting two nearest nodes turns out to be 424. The
characteristic path length $L$ is defined in terms of the network
distance $n_{ij}$, which represents the shortest path length
between nodes $i$ and $j$. Usually, the clustering coefficient $C$
also provides an important measure for a complex network. However,
since a few nodes of the subway system have only one nearest
neighbor, the clustering coefficient $C$ is not well defined.
Accordingly, we define the clustering coefficient $C^*$ of the
subway system, excluding the nodes which have one neighbor. The
eccentricity of node $i$ corresponds to the greatest distance
between $i$ and other node. The radius $R$ and the diameter $D$ of
a network are then defined to be the minimum eccentricity and the
maximum eccentricity, respectively, among all nodes. We further
define the efficiency according to
\begin{equation}
\epsilon \equiv
\frac{1}{N(N-1)}\sum_{i\ne j} \frac{1}{n_{ij}}. \label{eq:e}
\end{equation}
In the ideal case for the efficiency $\epsilon$, all nodes are
connected to each other, so that the network of $N$ nodes has
$N(N-1)/2$ links. Normalizing the efficiency to that in the ideal
case, we obtain the \textit{network efficiency} $E$, which takes a
value between zero and unity: $0\le E \le 1$.

We carry out measurement of the above quantities for the MSS
network and show the results in Table \ref{table:quantity}. Note
that unlike, e.g., an airline network, it is not conceivable for a
subway network to have all-to-all connections between stations.
The number of links in a real network of $N$ stations should be
far less than $N(N-1)2$, thus leading to the small value of the
network efficiency $E$. On the other hand, it is desirable to
build the network in such a way that the physical distance
measured along the links between a pair of stations is as short as
possible, compared with the actual distance (along the straight
line) between the two.  Accordingly, it is more appropriate to use
the physical distance $d_{ij}$ rather than the network distance
$n_{ij}$ in measuring the network efficiency $E$.  Whereas the
value of $E$ in the network distance represents the efficiency
with respect to the ideal but unrealistic network (with all-to-all
connections), the value in the physical distance measures the
efficiency with respect to the optimal network in reality.  In
terms of the physical distance, the characteristic path length,
diameter, and radius are given by $L=27.9$\,km, $D=139$\,km, and
$R=69.8$\,km, respectively.

\begin{figure}
\includegraphics[width=1.0\textwidth]{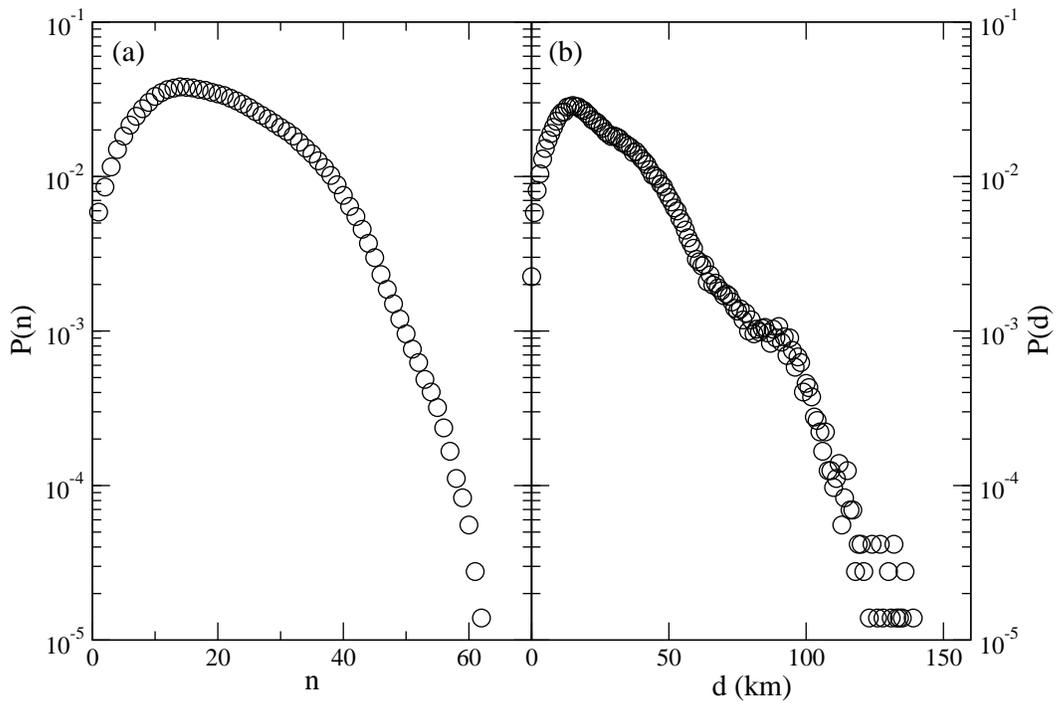}
\caption{\label{fig:shortest} Probability distribution of the
shortest path length in terms of (a) the network distance $n$ and
(b) the physical distance $d$ between stations on semi-log
scales.}
\end{figure}

Figure \ref{fig:shortest} exhibits the distribution of the
shortest path lengths, (a) in terms of the network distance (or
path length) and (b) in terms of the physical distance. Namely,
the horizontal axis of Fig. \ref{fig:shortest}(a) represents the
number of links between a pair of nodes (stations) while that of
(b) the physical distance between them.

We next investigate the passenger flows on the system. The MSS
network operates a smart card system which keeps track of the
travel information of every passenger.
Here we analyze the passenger flows on a single day, based on the
transaction data of the smart card on 24 June, 2005. The total
number of transactions or passenger flows in the MSS network was
as many as 4,909,316 on that specific day.

\begin{figure}
\includegraphics[width=1.0\textwidth]{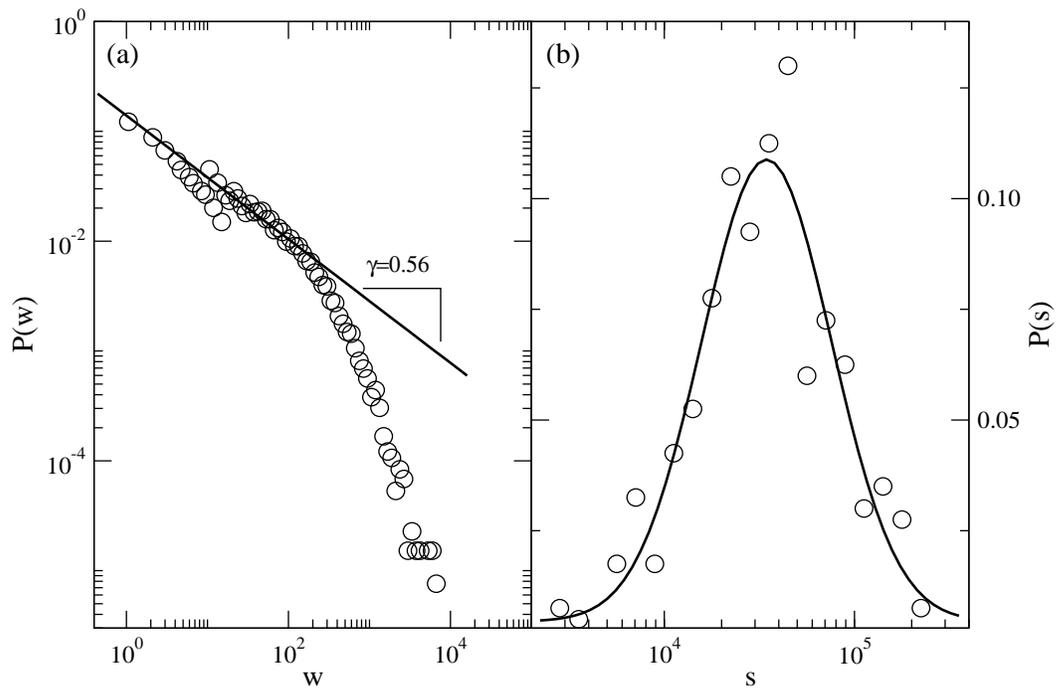}
\caption{\label{fig:weight} Probability distribution of (a) weight
$w$ on the log-log scale and (b) strength $s$ on the semi-log
scale. Lines are guides to the eye. In (a) the line has the slope
$-0.56$ whereas the line in (b) represents a log-normal
distribution.}
\end{figure}

The weight $w_{ij}$ of a link between stations $i$ and $j$ is
taken to be the sum of passenger flows in both directions on the
link, i.e., $i\rightarrow j$ and $j\rightarrow i$. The strength
$s_i$ of station $i$ is then defined to be the sum: $s_i \equiv
\sum_{j=1}^N w_{ij}$. Displayed in Fig. \ref{fig:weight} are the
obtained distributions of (a) weights and (b) strengths for the
MSS network. It is observed that the weight distribution $P(w)$
apparently exhibits power-law behavior with an exponent around
$0.56$, albeit restricted due to the finite system size;
in contrast, the strength distribution follows a log-normal
function with a peak at $s\approx 4\times 10^{4}$. Note that in
the subway network the weight of a link connecting two stations
represents the passenger flow between them and the strength of a
station corresponds to the number of passengers arriving at and
departing from that station. Accordingly, while passenger flows do
not have a characteristic size, numbers of passengers at single
stations do have. In a metropolis, most facilities are located
near stations, so that each station is naturally abundant in
passengers, the number of which reflects the capacity of
facilities located near the station. The fact that a majority of
stations are used by a similar number of passengers, corresponding
to the peak of the strength distribution, thus indicates that with
residential and commercial facilities taken into account, places
near most stations are already developed fully to accommodate
dense and compact location of facilities. The peak of the strength
distribution thus gives a measure for the characteristic capacity
of the facilities near a station in the fully urbanized Seoul.

\begin{figure}
\includegraphics[width=1.0\textwidth]{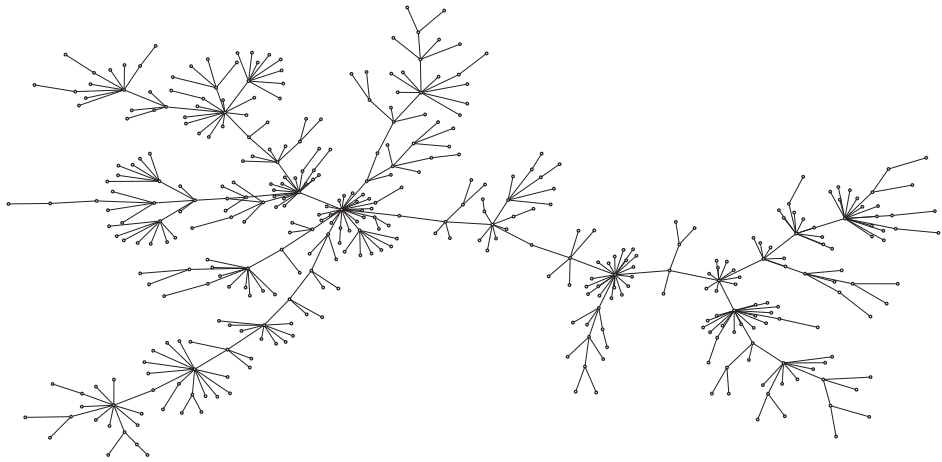}
\caption{\label{fig:mst} Maximum spanning tree of passenger flows,
consisting 380 stations in the Metropolitan Seoul Subway system.}
\end{figure}

\begin{figure}
\includegraphics[width=1.0\textwidth]{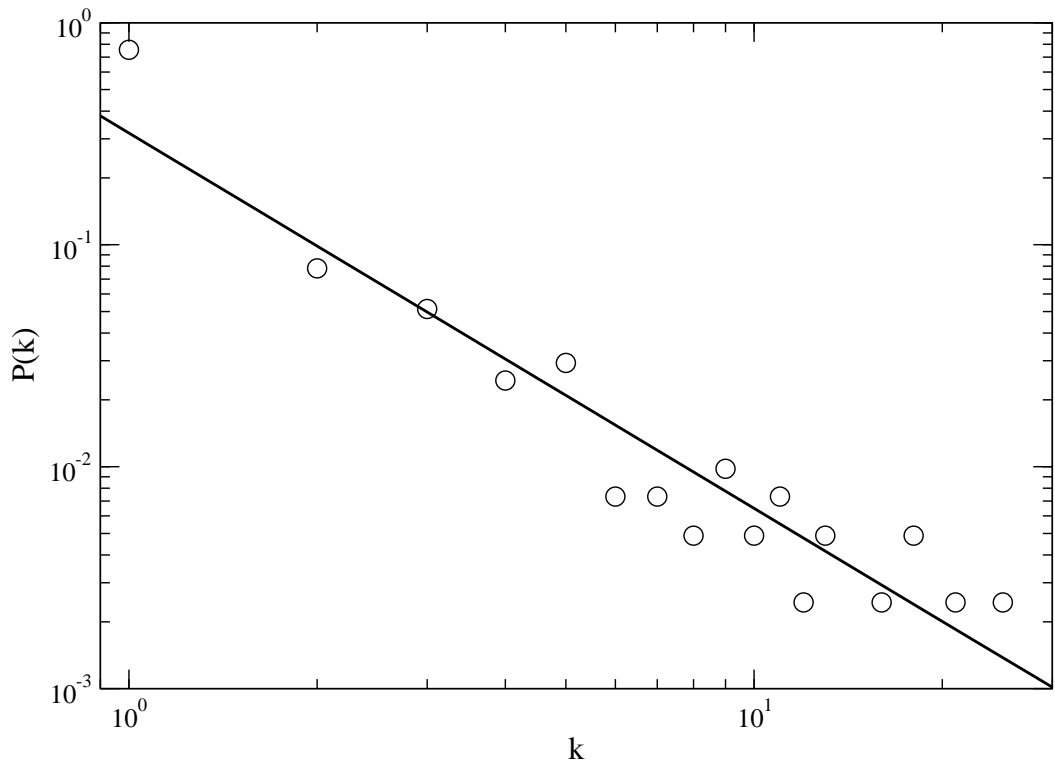}
\caption{\label{fig:degree} Probability distribution of the degree
$k$ of the maximum spanning tree in Fig. \ref{fig:mst} on the
log-log scale. The slope of the line, serving as a guide to the
eye, is given by $-1.7$.}
\end{figure}

Spanning trees are widely used to analyze a complex network
\cite{mantegna99}. Of particular interest is the minimum spanning
tree, which is constructed with weights not larger than those of
other possible spanning trees. In the subway system, on the other
hand, the link which has a larger passenger flow than others is
important, demanding to consider the {\it maximum} spanning tree.
Figure \ref{fig:mst} shows the maximum spanning tree of passenger
flows in the MSS system. We then compute the degree distribution
$P(k)$ of the maximum spanning tree, which is shown in Fig.
\ref{fig:degree}. It is observed to be consistent with a power-law
behavior: $P(k)\sim k^{-\gamma}$ with exponent $\gamma \approx
1.7$, obtained from the least-square fit.

To summarize, we have analyzed the Metropolitan Seoul Subway
system consisting of 380 stations, and obtained various network
measurements including the path length, clustering coefficient,
diameter, and radius as well as the efficiency of the network. The
path length, diameter, and radius have also been computed in terms
of the physical distance between stations. We have further
investigated the passenger flows in the system, and constructed
the maximum spanning tree of the flows. It is found that the
weight distribution displays a power-law behavior whereas the
strength distribution follows a log-normal one. Also revealed is
the power-law behavior of the degree distribution of the spanning
tree. The detailed analysis and implications are left for further
study.

\begin{ack}
This work was supported by the Korea Research Foundation through
Grant No. KRF-2006-B00022 and by the Sungshin Women's University
Research Grant of 2008.
\end{ack}

\newpage
\clearpage
\begin{table}
\begin{center}
\begin{tabular}{ccc}
\hline
&network&physical\\
&distance&distance\\
\hline
$N$&380&$\leftarrow$\\
$L$&$20.0$&$27.9$\,km\\
$n_\textrm{max}$&62&$139$\,km\\
$C^*$&$6.41\times10^{-3}$&$\leftarrow$\\
$D$&62&$139$\,km\\
$R$&31&$69.8$\,km\\
$E$&$7.86\times10^{-2}$&0.747\\
\hline
\end{tabular}
\caption{\label{table:quantity} Statistical properties of the
Metropolitan Seoul Subway network.}
\end{center}
\end{table}

\newpage
\clearpage

\newpage
\clearpage

\newpage
\clearpage

\newpage
\clearpage

\end{document}